\documentclass[a4paper,11pt,twocolumn,twoside]{article}
\pdfoutput=1
\usepackage{graphicx}
\usepackage[applemac]{inputenc}
\usepackage[all]{xy}
\usepackage{amsmath}
\newcommand{\captionfonts}{\small}

\makeatletter  % Allow the use of @ in command names
\long\def\@makecaption#1#2{%
  \vskip\abovecaptionskip
  \sbox\@tempboxa{{\captionfonts #1: #2}}%
  \ifdim \wd\@tempboxa >\hsize
    {\captionfonts #1: #2\par}
  \else
    \hbox to\hsize{\hfil\box\@tempboxa\hfil}%
  \fi
  \vskip\belowcaptionskip}
\makeatother   % Cancel the effect of \makeatletter

\title{How to cope with climate's complexity?}
\author{Michel Crucifix \\ \normalsize 
        Institut d'Astronomie et de Géophysique G. Lemaître \\ Université catholique de Louvain \\ 2 chemin du Cyclotron BE-1348 Louvain-la-Neuve, Belgium \\
\texttt{michel.Crucifix@uclouvain.be}}
\begin{document}
\onecolumn
\maketitle
% \linenumbers
\bibliographystyle{euroreview}
\begin{abstract}
Climate exhibits a vast range of dissipative structures. Some have characteristic times of a few days; others evolve on thousands of years. All these structures are interdependent; in other words,  they communicate. 
It is often considered that the only way to cope with climate complexity is to integrate the equations of atmospheric and oceanic motion with the finer possible mesh. Is this the sole strategy? Aren't we missing another characteristic of the climate system: its ability to destroy and generate information at the macroscopic scale? Paleoclimatologists consider that much of this information is present in palaeoclimate archives. It is therefore natural to build climate models such as to get the most of these archives.
The strategy proposed here is based on Bayesian statistics and  low-order non-linear dynamical systems, in a modelling approach that explicitly includes the effects of uncertainties.
Its  practical interest is illustrated through the problem of the timing of the next great glaciation. Is glacial inception overdue, or do we need to wait for another 50,000 years before ice caps grow again? Our results indicate a glaciation inception in 50,000 years.  
\end{abstract}
\pagestyle{headings}
\twocolumn

\begin{flushright}
\pagenumbering{arabic}
\textit{L'analyse mathématique peut déduire des phéno\-mènes gé\-néraux et simples l'expres\-sion des lois de la nature;
mais l'appli\-cation spéciale de ces lois à des effets très-composés exige une longue suite d'observations exactes.} 
\footnote{ Mathematical analysis allows you to deduce nature's laws from general and simple phenomena; but applying these laws to highly composite effects requires a long series of exact observations} \\ Joseph Fourier (1768 -- 1830)
\end{flushright}

\section{In troduction}

This quote by Joseph Fourier appeared first in the ``discours pr\'eliminaire'' of the \textit{analytical theory of heat} \cite{Fourier22aa}. At a time when the reversible Newtonian equations were championed by Pierre-Simon Laplace  (1749 -- 1827) and Joseph Louis Lagrange (1736 -- 1813), the irreversible equations governing heat propagation constituted a genuine mental revolution. With this sentence, Fourier arguably sets the foundations of complex system theory. He repeated it at least once, to conclude his \textit{mémoire sur les temp\'eratures du globe terrestre et des espaces planétaires} \cite{Fourier90aa} in which Fourier formulates what is known today as the ``greenhouse effect''. Fourier confesses that ``the question of Earth's temperature is one of the most important and difficult of all the Natural Philosophy''\cite{Fourier90aa}  and solving it was one central motivation for the theory of heat. Clearly, Fourier had fully perceived the complex character of the climate system. How, two centuries later, do we cope with climate's complexity? Which \textit{mathematical analysis} is the most appropriate to get the best out of \textit{observations}? With this paper we would like to  convince the reader that the most complex model is not necessarily the most useful. Predicting and understanding the climate system requires a consistency between the level of complexity of of observations, model prediction and what one wants to predict. Choosing the right model is thus also a question of information theory.

The case will be illustrated through a polemic currently taking place in the circle of Quaternary climate scientists. Here is it. As we shall see in more detail, the climate history of the past few million years is characterised by repeated transitions between ``cold'' (glacial) and warm (interglacial) climates. The first modern men were hunting mammoth during the last glacial era. This era culminated around 19,000 years ago \cite{lambeck00} and then declined rapildy. By 9,000 years ago climate was close to the modern one. The current interglacial, called the Holocene, has lasted long enough compared to previous interglacials. The polemic is about when it is supposed give way to a new glacial inception, keeping aside human activities that have most probably perturbed natural cycles.  

On the one side, 
Professor of Environmental Sciences Bill Ruddiman carefully inspected and compared palaeo-environmental information about the different interglacial periods. This comparison exercise let him to conclude that glacial inception is largely overdue \cite{ruddiman03,ruddiman07rge}. According to him, the Holocene was not supposed to be that long, but the natural glacial inception process was stopped by an anthropogenic perturbation that began as early as 6,000 years ago (rice plantations and land management by antique civilisations). On the other side, Professor André Berger and colleagues developed a mathematical model of the climate system, rated today as a ``model of intermediate complexity''\cite{gallee91,gallee92} including 15,000 lines of FORTRAN code to solve the dynamics of the atmosphere and ice sheets on a spatial grid of 19 x 5 elements, with a reasonably extensive treatment of the shortwave and longwave radiative transfers in the atmosphere. Simulations with this model led Berger and Loutre conclude that glacial inception is not due before 50,000 years as long as the CO$_2$ atmospheric concentration stays above 220 ppmv. \cite{Berger2002An-exceptionall}
Who is right? Both (Crucifix and Berger argued that the two statements are not strictly incompatible \cite{crucifix06eos})? None? Both Ruddiman and Berger judge that it is possible to predict climate thousands of years ahead but is it a realistic expectation after all? Michael Ghil wondered ``what can we predict beyond one week, for how long and by what methods?'' in a paper entitled ``Hilbert's problem of the geosciences in the XXIst century'' \cite{ghil01hilbert}. This is the fundamental motiviation behing the present article.

\section{Steps towards a dynamical model of palaeoclimates}
\subsection{Some general remarks about complex system modelling \label{remarks}}

%The proper of complex systems --- and climate is definitely one --- is that it is impossible to know them fully. It is impossible to know precisely the position and size of any molecule of air and ocean and to know all the chemical reactions occurring at any given time. 
%In spite of this ignorance, it is possible to make useful predictions about the evolution of a complex system by taking advantage of the existence of organised patterns at a scale much beyond that of the molecule.  These patterns constitute information that is susceptible of being modelled, understood and predicted.

A system as complex as climate is organised at different levels : clouds, cloud systems, synoptic waves, planetary waves, pluri-annual oscillations such as El-Niño, glacial-interglacial cycles\ldots.  
It is not our purpose to explain here how, in general, patterns emerge in complex systems (\cite{Haken06aa,Nicolis07aa} are up-to-date references on the subject) but it is useful to have a few notions in mind. Complex systems and their components act as information processors. This means that their dynamics is such that they can destroy, amplify and even create  information. The difficult mental barrier to overcome for physicists accustomed to Newtonian mechanics is that while the definition of information is subjective (it depends on a \textit{choice} of variables describing the system), the processes of destruction and creation of information rely on general theories.

At the risk of being schematic, one may say information is created by instabilities (necessarily fed by some source of energy), and it is destroyed by relaxation processes (return to equilibrium).  The resulting stationary patterns are a balance between both. A typical laboratory example is the B\'enard Cells.

In the atmosphere, local hydrodynamical instabilities result in planetary waves, such as the ones responsible for dominant north-westerly winds in Canada and south-westerly winds in Europe. 
A pure linear thinker might estimate that the hypothetical butterfly that caused the initial atmospheric instability 
is the cause of the wave. It is, indeed, chronologically the first of a sequence of events that lead to the macroscopic pattern. On the other hand, the ``non-linear'' thinker will observe that the macroscopic properties of the wave (for example,  its spectral characteristics)
do not depend on the position and time of the butterfly that triggered the initial instability. This information has been destroyed by the system dynamics. In this view, the ``causes'' of the wave are the conditions that made the initial instability possible. 

This lead us to conclude that although there is no way to know the climate system fully (it is impossible to know precisely the position and size of any molecule of air and ocean and to know all the chemical reactions occurring at any given time) it is possible to make useful predictions about the evolution of some macroscopic variables by taking advantage of organised patterns.  
It is therefore sensible to define climate's state using variables relevant for what one wants to predict. Our goal is to predict the next glacial inception, so we will concentrate on variables such as ice volume and carbon dioxide concentration. These are called  order parameters. They describe system's state, but incompletely so, and our goal is to establish balance equations for these variables (cf. \cite{Nicolis07aa}, pp. 36-37). 

Behind the mere desire of predicting the next glacial inception, our more general ambition is to \textit{identify} and \textit{understand} which constraints mostly determine climate evolution at the glacial-interglacial time scale. What do we need to know to predict the next glacial inception, and why is this information important ? This question prompts us to build models that take explicitly into account our knowledge and associated  uncertainties, and \textit{validate} these models, that is, to test that the underlying assumptions are compatible with observations (e.g.: \cite{Rougier2007Probabilistic-i}).

We have already seen that instabilities are information generators. Macroscopic patterns therefore depend on the parameters that control the growth of such instabilities. Only in relatively idealised and simple cases do we know these parameters with enough accuracy to correctly predict the macroscopic order parameters. In most natural cases, instabilities are so numerous and intricate that the resulting effects cannot possibly be predicted without appropriate observations.
Namely, Saltzman repeatedly insisted \cite{saltzman84aa,saltzman02book} on the fact that neither current observations nor modelling of the present state of the atmosphere can possibly inform us of the ice-sheet mass balance with sufficient accuracy to predict their evolution at the timescale of several thousands of years. We need to look at palaeoclimate history to get this information.

The relevant strategy strategy will therefore consist in using both first principles and empirical information to formulate the balance equations governing  the dynamics of glacial-interglacial cycles. Of course, these  equations must be \textit{compatible} with our knowledge of atmosphere and ocean dynamics at the interannual time scale, but we accept the fact that we do not immediately deduce them from it. The process by which model parameters are estimated on the basis of observations is called \textit{calibration} \cite{Rougier2007Probabilistic-i}.

This empirical, or inductive approach is acceptable as long as it respects the fundamental statements of information theory. In particular, any system of time-differential equations that reproducibly predicts the evolution of macroscopic variables must be dissipative (the volume of initial conditions must collapse to an attractor) (\cite{Jaynes79aa} and \cite{Nicolis07aa}, pp. 195 onwards. )

\subsection{Empirical evidence about the Quaternary\label{emperical}}
Building a robust theory of glacial-interglacial cycles  requires a profound knowledge of the Quaternary. This section is intended to provide a glimpse at the vast amount of knowledge that scientists have accumulated on that period before we propose a mathematical methodology to address Ruddiman's hypothesis. 

\subsubsection{The natural archives}
By the nineteen-twenties, geomorphologists were able to correctly interpret the glacial moraines and alluvial terraces as the left-overs of previous glacial inceptions. Penk and Brückner (\cite{Penck09aa}, cited by \cite{berger88}) recognised four previous glacial e\-pochs, named the Günz, Mindel, Riss and Würm. 
The wealth of data on the Quaternary environments that has since been collected and analysed by field scientists can be appreciated from the impressive four-volume encyclopedia of Quaternary Sciences recently edited by Elias \cite{elias07encyclop}. Analysis and interpreting palaeoenvironmental data involve a huge variety of scientific disciplines, including geochemistry, vulcanology, palaeobiology, nuclear physics, stratigraphy, sedimentology, glacial geology and ice-stream modelling.

Only a schematic overview of this rich and intense field of scientific activity could possibly be given here. The reader will find most of the relevant references in the encyclopedia, and only a few historical ones are provided here.

\textbf{Stable isotopes} constitute one important class of natural archives.  
It is indeed known since the works of Urey \cite{Urey48aa}, Buchanan \cite{Buchanan53aa} and Dansgaard \cite{dansgaard64} that physical and chemical transformations involved in the cycles of water  and carbon fractionate the isotopic composition of these elements. 
To take but a few examples, ice-sheet water is depleted in oxygen-18 and deuterium compared to sea water; clouds formed at low temperatures are more depleted in oxygen-18 and deuterium than clouds formed at higher temperatures; organic matter is depleted in $^{13}$C, such that inorganic carbon present in biologically active seas and soils is enriched in $^{13}$C. $^{15}N$  is another useful stable palaeo-environmental indicator sensitive to the biological activity of soils.
  The isotopic compositions of  water and biogenic carbon are extracted deep-sea sediments,  ice and air trapped in ice bubbles, palaeosols, lake-sediments and stalagmites. One of the first continuous deep-sea record of glacial-interglacial cycles was published by Cesare Emiliani \cite{emiliani55}.

  \textbf{Radioactive tracers} are used to estimate the age of the record and rate of ocean water renewal. At the timescale of the Quaternary, useful mother-daughter pairs are $^{230}$Th / $^{238,234}$U (dating carbonates), and $^{40}$K / $^{40}$Ar in potassium-bearing minerals. The ratio 
  $^{230}$Th / $^{231}$Pa is a useful indicator of ocean circulation rates. 

  \textbf{The chemical composition of fossils} is also indicative of past environmental conditions. In the ocean, cadmium, lithium, barium and zinc trapped in the calcite shells of foraminifera indicate the amount of nutrients at the time of calcite formation, while the foraminifera content in magnesium and strontium are empirically correlated to water-temperature. 

  Glaciologists have also developed ambitious programmes to analyse \textbf{the composition of air} (oxygen, nitrogen, plus trace gases such as methane, carbon-dioxide and nitrogen oxide, argon and xenon) trapped in ice accumulating on ice sheets, of which the \textit{European Project for Ice Core in Antarctica} is a particularly spectacular achievement\cite{jouzel07science}.  It was demonstrated that the central plateaus of Antarctica offer a sufficiently stable environment to reliably preserve air's chemical composition over several hundreds of thousands of years. The chemical composition of water is sensitive to atmospheric circulation patterns and sea-ice area.

  Additional sources of informations are obtained from a variety of marine and continental sources. Plant and animal fossils (including pollens) trapped in lakes, peat-bogs, palaeosols and marine sediments provide precious indications on the palaeoenvironmental conditions that conditioned their growth. Their presence (quantified by statistical counts) or absence may be interpreted quantitatively to produce palaeoclimatic maps \cite{climap81}. Preservation indicators of  ocean calcite fossils are used to reconstruct   the history of ocean alkalinity.  Palaeosols and wind-blown sediments (loess) provide precious indications on past aridity at low-latitudes. The loess grain-size distribution is also sensitive to atmospheric circulation patterns. Geomorphological elements remain a premium source of information about the configuration of past ice sheets, which is complemented by datable evidence (typically coral fossils) on sea-level. 
  \subsubsection{The structure of Quaternary climate changes}

  It is barely straightforward to appreciate which fraction of the information available in a climate record is relevant to understand climate dynamics at the global scale. For example, minor shifts in oceanic currents may have sensible effects on the local isotopic composition of water with however no serious consequence for glacial-interglacial cycle dynamics. One strategy is to collect samples from many areas of the world and average them out according to a process called ``stacking''. One of the first ``stacks'', still used today, was published by John Imbrie and colleagues \cite{imbrie84} in the framework of the \textit{Mapping spectral variability in global climate project}. It is usually referred to as the \textsc{Specmap} stack. Here we concentrate on the more recent compilation provided by Lisiecki and Raymo \cite{lisiecki05lr04}, called LR04. The stack was obtained by superimposing 57 records of the oxygen-18 composition of benthic foraminifera shells. Benthic foraminifera live in the deep ocean and therefore record the isotopic composition of deep water (an indicator of past ice volume). However, there is an additional fractionation associated to the calcification process, which is proportional to water temperature. The isotopic composition of calcite oxygen is reported by a value, named $\delta^{18}$O$_c$, giving the relative enrichment of oxygen-18 versus oxygen-16 compared to an international standard.  High $\delta^{18}$O indicates either low continental ice volume and / or high water-temperature.

\begin{figure}[t]
\begin{center}
\includegraphics[width=\columnwidth]{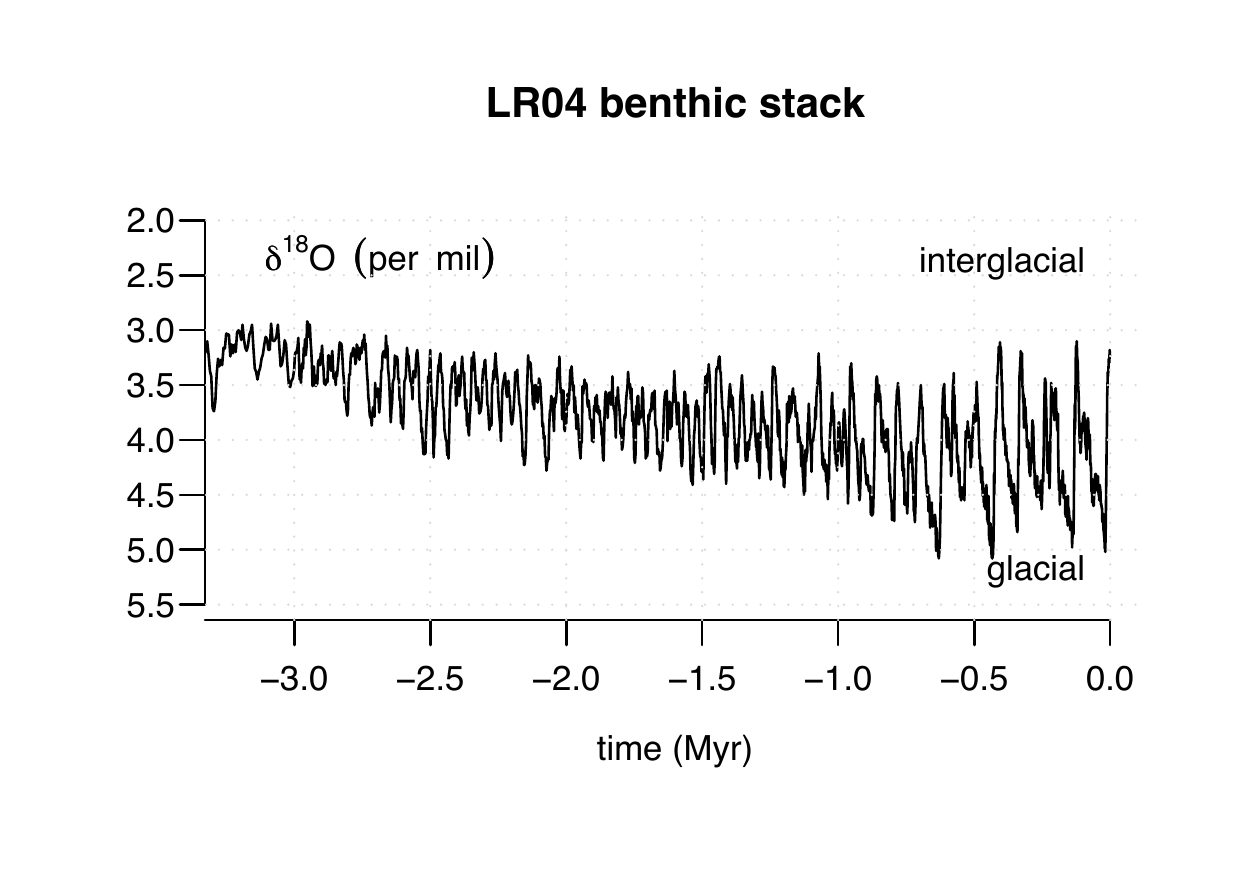}
\end{center}
\caption{The LR04 benthic $\delta^{18}$O stack constructed by the graphic correlation of 57 globally distributed benthic $\delta^{18}$O records \cite{lisiecki05lr04}. Note that the full stack goes back in time to -5.2 Myr  (1 Myr = 1 million years). The signal is the combination of global ice volume (low $\delta^{18}$O corresponding to low ice volume) and  water temperature (low $\delta^{18}$O corresponding to high temperature). The Y-axis is reversed as standard practice to get ``cold'' climates down. Data downloaded from \texttt{www.lorraine-lisiecki.com}.}
\label{fig:LR04}
\end{figure}

Visual inspection of the LR04 stack (Figure \ref{fig:LR04}) nicely evidences the gradual transition from the Pliocene --- warm and fairly stable --- to the spectacular oscillations of the late Pleistocene. The globally averaged temperature at the early Pliocene was about 5$^\circ$ C higher than today (\cite{Raymo96aa} and references therein); that one at the last glacial maximum (20,000 years ago) was roughly 5$^\circ$ C lower. The central research question we are busy with is to characterise these oscillations, understand their origin and qualify their predictability. 

\begin{figure}[t]
\begin{center}
\includegraphics[width=\columnwidth]{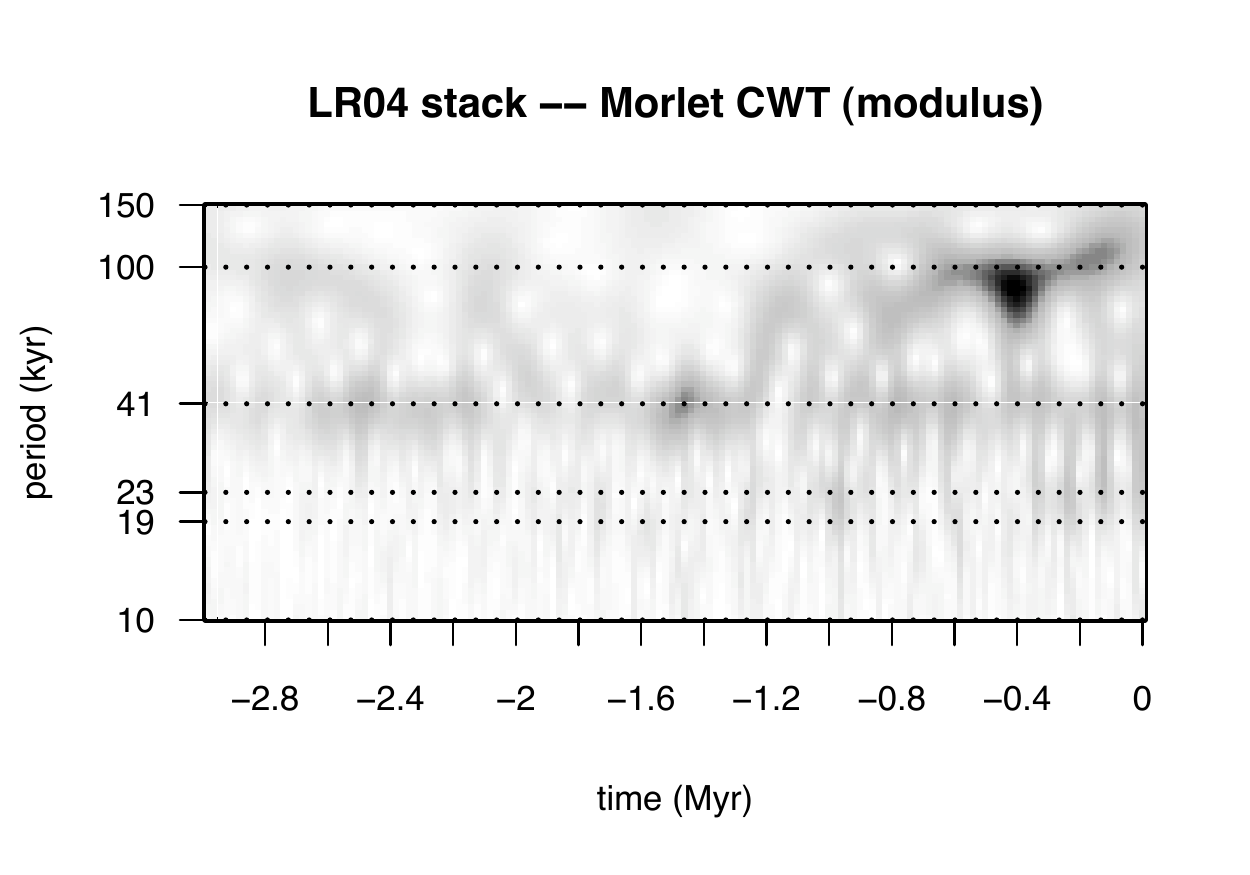}
\end{center}
\caption{Modulus of the continuous Morlet Transform of the LR04 stack according to the algorithm given in Torrence and Compo \cite{torrence98} using $\omega_0=5.4$ but with a normalisation $c(s)=s/\sqrt{\Delta t}$. R routine adapted by J.L. Melice and the author from the original code supplied by S. Mallat \cite{Mallat98aa}. The wavelet transform evidenes the presence of quasi-periodic signals (shades) around periods of 41 kyr (1~kyr = 1,000 years) and 1000 kyr.}
\label{fig:LR04morlet}
\end{figure}

The Morlet Continuous wavelet transform provides us with a first outlook on the backbone of these oscillations (Figure \ref{fig:LR04morlet}). The LR04 record is dominated most of the time by a 40,000-yr signal until roughly 900,000 years ago, after which the 40,000-yr signal is still present but topped by longer cycles. At the very least, this picture should convince us that LR04 contains structured information susceptible of being modelled and possibly predicted.

How many differential equations will be needed? There will be no clear-cut answer to that question. Time-series extracted from complex systems are sometimes characterised by their \textit{correlation dimension}, which is an estimator for the fractal dimension of the corresponding attractor \cite{GRASSBERGER83aa}. The first estimates for the Pleistocene were provided by Nicolis and Nicolis \cite{nicolis86} ($d=3.4$) and Maasch et al.\cite{Maasch89aa} ($4\le d \le 6$). For this article we calculated correlation dimension estimates for the LR04 stack ($d=1.54$) and the HW04 stack \cite{huybers04Pleistocene} ($d=3.56$). HW04 is  similar to LR04 but it is based on different records and dating assumptions. Several authors, including Grassberger himself \cite{GRASSBERGER86aa,Pestiaux84aa,VAUTARD89aa} have discouraged the use of correlation dimension estimates for the ``noisy and short'' time series typical of the Quaternary because they are  overly sensitive to sampling and record length. They are therefore unreliable.%and they concluded that it is not possible to reliably distinguish stochastic from low-order chaotic noise in such series.

\begin{figure*}[t]
\includegraphics[width=0.7\textwidth]{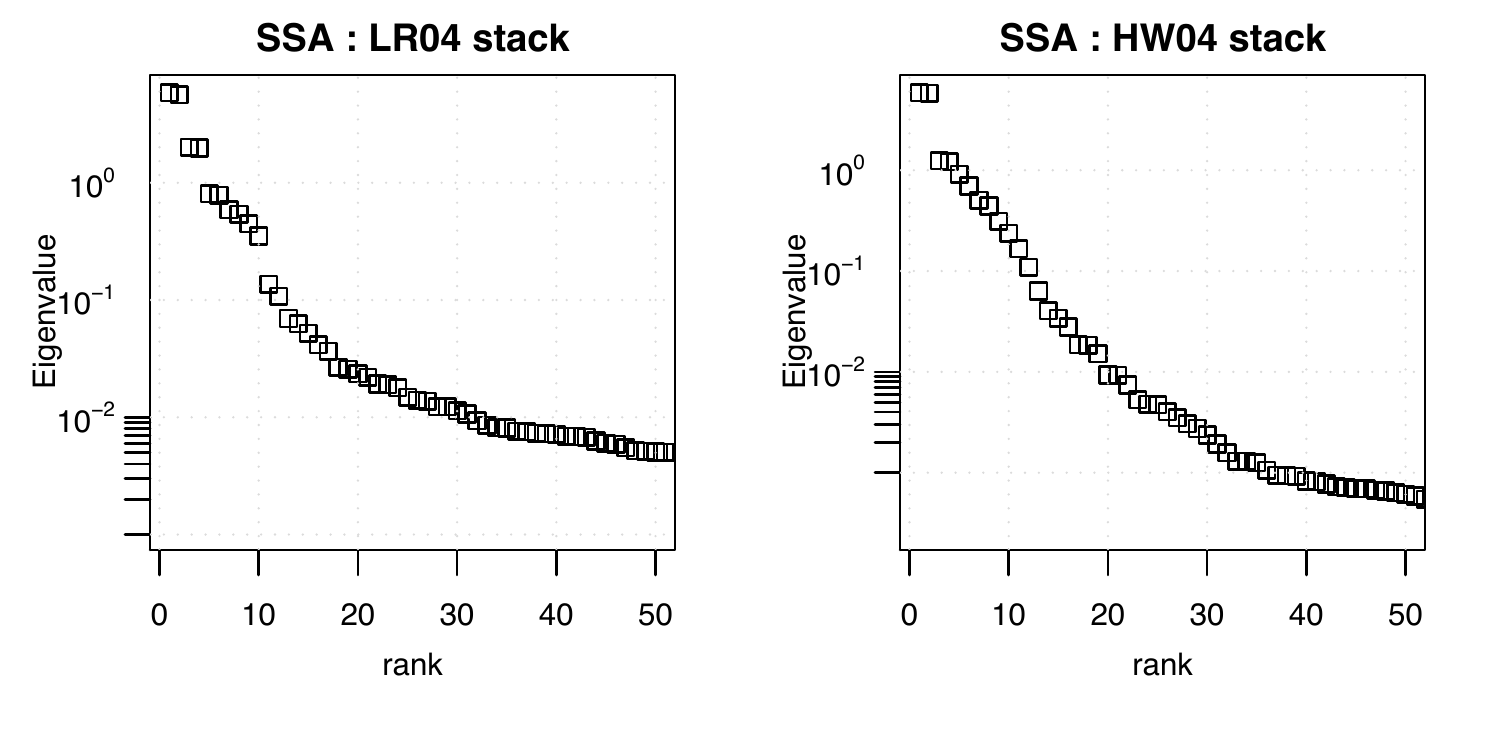}
\caption{Single Spectrum Analysis  (SSA) of the LR04 and HW04 benthic stacks. Displayed are the eigenvalues of the lagged-covariance matrix of rank $M=100$ as given by \cite{Ghil02aa}, equation (6).  The records were cubic-spline interpolated ($\Delta t = 1 kyr$) and only the most recent 900 kyr were kept. The SSA decomposition of LR04 is very typical:  it evidences three oscillators (recognisable as pairs of eigenvectors), then about four modes that are generally interpreted as harmonics of the dominant ones, and finally a number of modes typically interpreted as stochastic background. The HW04 stack contrasts with LR04 because the dominant modes are not so easily evidenced.  HW04 uses less benthic records than LR04, but it also relies on more conservative dating assumptions and this probably resulted in blurring the quasi-periodic components of the signal. HW04 data were obtained from \texttt{www.people.fas.harvard.edu/~phuybers/}.}
\label{fig:LR04ssa}
\end{figure*}

In response to this problem Ghil and colleagues  \cite{VAUTARD89aa,YIOU94aa}
promoted single-spectrum analysis, in which a time series is linearly decomposed into a number of prominent modes (which need not be harmonic), plus a number of small-amplitude modes. Assuming  that the two groups are indeed separated by an amplitude gap, the first group provides the low-order backbone of the signal dynamics while the second group is interpreted as stochastic noise. Single spectrum analysis was applied with a certain success to various sediment and ice-core records of the few last-glacial interglacial cycles \cite{YIOU94aa} and have in general confirmed that the backbone of climate oscillations may be captured as a linear combination of a small number of amplitude and / or frequency-modulated oscillations. Single-spectrum analysis of the last million years of LR04 (Figure \ref{fig:LR04ssa}) confirms this statement.

\subsubsection{The Achille heel}
Now time has come to mention a particularly difficult and intricate issue: dating uncertainty in palaeoclimate records. No palaeoclimate record is dated with absolute confidence. Marine sediments are coarsely dated by identification of a number of reversals of Earth's magnetic field, which have been previously dated in rocks by radiometric means (\cite{raymo97} and references therein). Magnetic reversals are pretty rare (four of them over the last 3 million years) and their age is known with a precision no better than 5,000 years. Local sedimentation rates may vary considerably between these time markers such that any individual event observed in any core taken in isolation is hard to date.  Irregularities in the sedimentation rate blur and destroy information that might otherwise be evidenced by spectral analysis. 

One strategy to contend this issue is to assume synchrony between oscillation patterns identified in different cores. Statistical tests may then be developed on the basis that dating errors of the different cores are independent. For example, Huybers (2007) \cite{huybers07obl} considered the null-hypothesis that glacial-interglacial transitions (they are called \textit{terminations} in the jargon of palaeoclimatologists) are independent on the phase of Earth's obliquity. While this null-hypothesis could not  be rejected on the basis of a single record, the combination of 14 cores allowed him to reject it with 99\% confidence, proving once more the effect of the astronomical forcing on climate.  First tests of this kind were carried out by Hays et al. in a seminal paper \cite{hays76}.
Note that in many cases the oscillation patterns recognised in different cores are so similar that it is hard to dispute the idea of somehow ``matching them'', but it is remarkable that rigorous statistical tests assessing the significance of a correlation between two ill-dated palaeoclimate records are only being developed (Haam and Huybers, manuscript in preparation).

Another strategy is known as \textit{orbital tuning}. 
The method consists in 
squeezing  or stretching the time-axis of the record to match the evolution of one or a combination of orbital elements, possibly pre-filtered by a climate model  \cite{hays76,imbrie84}. The method undeniably engendered important and useful results (e.g. \cite{SHACKLETON90aa}), but the astute reader has already perceived its potential perversity: orbital tuning injects a presumed link between orbital forcing and the record.  Experienced investigators recognise that orbital tuning has somehow contaminated most of most of the dated palaeoclimate records available in public databases.  This has increased the risk of tautological reasoning. 

For example, compare the two SSA analyses shown in figure \ref{fig:LR04ssa}. As we mentioned, LR04 and HW04 are two stacks of the Pleistocene but LR04 contains more information. It is made of more records (57 instead of 21 in HW04) and it is astronomically tuned. We can see from the SSA analysis that LR04 presents more quasi-periodic structures than HW04 (recall that quasi-periodic modes are identified as pairs of eigenvalues with quasi the same amplitude). Why is this the case? Is this because age errors in HW04 blurred the interesting information, or is it because this information has been artificially injected in LR04 by the tuning process? There is probably a bit of both (but note that HW04 displays a similar wavelet structure as LR04).

\begin{figure*}[t]
\includegraphics[width=\textwidth]{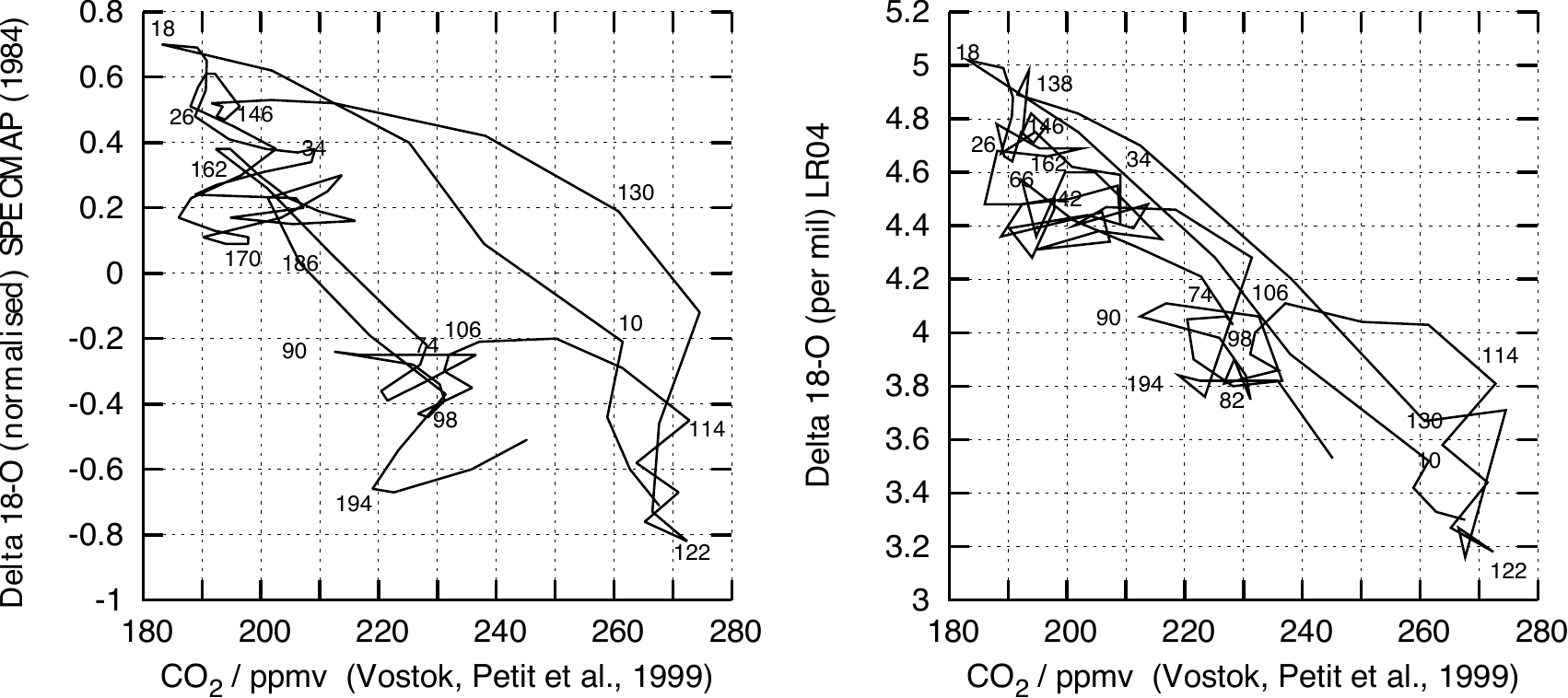}
\caption{The concentration in CO$_2$ measured in the Vostok ice core record \cite{petit99} over the last glacial-interglacial cycle is plotted versus two proxies of continental ice volume: \textit{(left)} : The planctonic $\delta^{18}$O stack by Imbrie et al. (1984) and \textit{(right)} : The benthic $\delta^{18}$O stack by Lisiecki and Raymo (2004). Numbers are dates, expressed in kyr BP (before present).  While the Imbrie stack suggests an hysteresis behaviour with CO$_2$ leading ice-volume variations, the picture based on LR04 is not so obvious.}
\label{fig:phase}
\end{figure*}

Leads and lags between CO$_2$ and ice volume is another difficult problem where risks posed by hidden dating assumptions and circular reasoning lie at every corner. 
Here is one typical illustration: Saltzman and Verbitsky
 showed at several occasions (e.g.:  \cite{SALTZMAN94ab})
 a phase diagram showing the SPECMAP  $\delta^{18}$O stack versus the first full ice-core records of CO$_2$ available at that time \cite{barnola87,jouzel93}. It is reproduced here (Figure \ref{fig:phase}, left). The phase diagram clearly suggests that CO$_2$ leads ice volume at the 100-kyr time scale. However, a detailed inspection of the original publications reveals that the SPECMAP record was astronomically tuned, and that the Vostok time-scale uses a conventional date of isotopic stage 5.4 of  110 kyr BP \ldots by reference to SPECMAP \cite{jouzel93} ! The hysteresis is therefore partly conditioned by arbitrary choices.
On Figure  \ref{fig:phase} we further illustrate the fact that the shape of the hysteresis depends on the stack record itself. 
 The situation today is that there is no clear consensus about the phase relationship between ice volume and CO$_2$ at the glacial interglacial time scale (compare \cite{Kawamura07aa,Ruddiman03aa,shackleton00}). According to the quite careful analysis of \cite{Ruddiman03aa}, CO$_2$ leads ice volume at the precession (20 kyr) period, but CO$_2$ and ice volume are  roughly synchronous at the obliquity (40 kyr) period.  Current evidence about the latest termination is that decrease in ice volume and the rise in CO$_2$ were grossly simultaneous and began around 19,000 years ago \cite{Kawamura07aa,yokoyama00lgm}

\subsection{Getting physical laws into the model}
So far we learned that palaeoclimate oscillations are structured and that it is not unreasonable to attempt modelling them with a reduced order model forced by the astronomical variations of Earth's orbit. 
What is the nature of the physical principles to be embedded in such a model, and how can they be formalised? The history of Quaternary modelling is particularly enlightening in this respect (the reader will find in \cite{berger88} an extensively documented review of Quaternary climates modelling up to the mid-eighties).
After Joseph Adhémar (1797 -- 1862) \cite{adhemar1842} suggested that the cause of glaciations is the precession of the equinoxes, Joseph John Murphy and James Croll (1821 -- 1890) argued about how precession may affect climate. Murphy maintained that cold summers (occurring when summer is at aphelion) favour glaciation \cite{Murphy76aa}, while Croll considered that cold winters are critical \cite{croll1875}.

Croll's book demonstrates a phenomenal encyclopaedic knowledge.  His judgements are at places particularly far-sighted, but they are barely substantiated by the \textit{mathematical analysis} Fourier was so much insistent about.  
The nature of his arguments are essentially phenomenological, if not at places frankly rhetorical. 

Milutin Milankovitch (1879 -- 1958) is then generally quoted as the one having most decisively crossed the step towards mathematical climatology. In a highly mathematical book that crowns a series of articles written between 1920 and 1941 \cite{milankovitch41}, Milankovitch extends Fourier's work to estimate the zonal distribution of Earth's temperature from incoming solar radiation. He also computes the effects of changes in precession, eccentricity and obliquity on incoming solar radiation at different latitudes to conclude, based on geological evidence, that summer insolation is indeed driving glacial-interglacial cycles.

Mathematical analysis is the process that allows Milankovitch to deduce the consequences of certain fundamental principles, such as the laws of Beer, Kirchhoff and Stefan, on global quantities such as  Earth's temperature. On the other hand, Milankovitch uses empirical macroscopic information, such as the present distribution of the snow-line altitude versus latitude, to estimate the effects of temperature changes on the snow cover. In today's language, one may say that Milankovitch had accepted that some informations cannot be immediately inferred from microscopic principles because they depend on the way the system as a whole has been dealing with its numerous and intricate constraints (Earth's rotation, topography, air composition etc.).

The marine-record study published by Hays, Imbrie and Shackleton \cite{hays76} is often cited as the most indisputable proof of Milankovitch's theory. Hays et al.  identified three peaks in the spectral estimate of climate variations that precisely correspond to the periods of obliquity (40 kyr) and precession (23 kyr and 19 kyr) calculated analytically by André Berger \footnote{the supporting papers by Berger would only appear in the two following years \cite{berger77,Berger77ber2,berger78}; Hays et al. based themselves on a numerical spectrum estimate of the orbital time-series provided by Vernekar \cite{vernekar72}}. 

However, \textit{sensu stricto}, Milankovitch's theory of ice ages was invalidated by evidence --- already available in an article by Broecker and van Donck \cite{broecker70}  --- that the glacial cycle is 100,000 years long, ice build up taking about 80,000 years and termination about 20,000 years \cite{broecker70,hays76}. Neither the 100,000-year duration of ice ages, nor their saw-tooth-shape were predicted by Milankovitch. 
The bit Milankovitch's theory is missing is the \textit{dynamical aspect} of climate's response. Glaciologist Weertman \cite{weertman76} consequently addressed the evolution of ice sheet size and volume by means of an ordinary differential equation, thereby opening the door to the use of dynamical system theory for understanding Quaternary oscillations. %The works led by Ghil (e.g.: \cite{ghil81}) and Saltzman \cite{saltzman02book} constitute important references in this respect.

In the meantime, general circulation models of the atmosphere and oceans  running on supercomputers became widely available (cf. \cite{randall00} for a review), and 
used for palaeoclimate purposes \cite{broccoli87lgm,kutzbach81,mitchell93}. The interest of these models is that they provide a consistent picture of the planetary dynamics of the atmosphere and the oceans. Just as Milankovitch applied Beer and Kirschoff's laws to infer Earth's temperature distribution, general circulation models allow us to deduce certain aspects of the global circulation from our knowledge of balance equations in each grid cell. 
However, these balance equations are uncertain and quantifying the consequences of these uncertainties  at the Earth global scale is a very deep problem that only begins to be systematically addressed \cite{allen00uncertainty}. While general circulation models  are undeniably useful to constrain the immediate atmospheric response to changes in orbital parameters, they are far too uncertain to reliably estimate glacial accumulation rates with enough accuracy to predict the evolution of ice sheets over tens of thousands of years \cite{saltzman84aa}. 

In the following sections we will concentrate on a 3-dimensional climate dynamical model written by Saltzman. This choice was guided by the ease of implementation as well as the impressive amount of supporting documentation \cite{saltzman02book}. 
However, they were numerous alternatives to this choice. The reader is referred to the article by Imbrie et al. \cite{Imbrie92aa} and pp. 264-265 of Saltzman's book \cite{saltzman02book} for an outlook with numerous references organised around the dynamical concepts proposed to explain glacial-interglacial cycles (linear models, with or without self-sustained oscillations, stochastic resonance,  model with large numbers of degrees of freedom).

The series of models published by Ghil and colleagues \cite{kallen1979,ghil81,letreut83} are among the ones having the richest dynamics. They present self-sustained oscillations with a relatively short period (6,000 years). The effects of the orbital forcing are taken into account  by means of a multiplicative coefficient in the ice mass balance equation. This causes non-linear resonance between the model dynamics and the orbital forcing.  The resultgin spectral response presents a rich background with  multiple harmonics and band-limited chaos. More recently, Gildor and Tziperman \cite{gildor01ab} proposed a model where sea-ice cover plays a central role. In this model, termination occurs when extensive sea-ice cover reduces ice accumulation over ice sheets. Like Saltzman's, this model presents 100-kyr self-sustained oscillations that can be phase-locked to the orbital forcing.

Field scientists with life-long field experience have also proposed models usually qualified as ``conceptual'', in the sense that  they are formulated as a worded causal chain inferred from a detailed inspection of palaeoclimate data without the support of differential equations. Good examples are \cite{Ruddiman03aa,Imbrie92aa,imbrie93,Ruddiman2006Ice-driven-CO2-}. In the two latter references, Ruddiman proposes a direct effect of precession on CO$_2$ concentration and tropical and southern-hemisphere sea-surface temperatures, while obliquity mainly affects the  hydrological cycle and the mass-balance of northern ice sheets. 
%This suggests that the traditional summer solstice insolation at 65 $^o$N used in the above dynamical systems on the basis of Milankovitch's works does not contain enough information to predict glacial-interglacial cycles. 

\subsection{The Saltzman model (SM91)}
As a student of Edward Lorenz, Barry Saltzman ( - 2002)  contributed to the formulation and study of the famous Lorenz63 dynamical system \cite{lorenz63} traditionally  quoted as the archetype of low-order chaotic system\footnote{The \textit{acknowledgments} of the Lorenz (1963) paper reads: ``The writer is indebted to Dr. Barry Saltzman for bringing to his attention the existence of nonperiodic solutions of the convection equations.}. Saltzman was therefore in an excellent position to appreciate the explanatory power of dynamical system theory. Between 1982  and 2002 he and his students published a small dozen of dynamical systems deemed to capture and explain the dynamics of Quaternary oscillations \cite{Saltzman82aa,saltzman84aa,saltzman90sm,saltzman91sm,saltzman93,saltzman02book}. In the present  article we choose to analyse the ``palaeoclimate dynamical model'' published by Saltzman and Maasch (1991) \cite{saltzman91sm}.  We will refer to this model as SM91.

Saltzman estimated that the essence of Quaternary dynamics should be captured by a three-degree-of-freedom dynamical system, possibly forced by the variations in insolation caused by the changes in orbital elements \cite{saltzman84aa}. The evolution of climate at these time scales is therefore represented by a trajectory in a 3-dimensional manifold, which Saltzman called the ``central manifold''. The three variables are ice volume ($I$), atmospheric CO$_2$ concentration ($\mu$) and deep-ocean temperature ($\theta$). It is important to realise that Saltzman did not ignore the existence of climate dynamics at shorter and longer time scales than those that characterise the central manifold, but he formulated the \textit{hypothesis} that these modes of variability may be represented by \textit{distinct} dynamical systems. 
In this approach, the fast relaxing modes of the complex climate system are in thermal equilibrium with its slow and unstable dynamical modes. This assumption is called 
the ``slaving principle'' and it was introduced by Haken \cite{Haken81aa}.

The justification of time-scale decoupling is a very delicate one and it deserves a small digression.
In some dynamical systems, even small scale features may truly be informative to predict large-scale dynamics. This phenomenon, called ``long-range interaction'', happens in the Lorenz63 model \cite{NICOLIS95aa}. The consequence is  that one might effectively ignore crucial information by averaging the fast modes and simply assume that they are in thermal equilibrium. To justify his model, Saltzman used the fact that there is a ``spectral gap'', that is a range of periods with relatively little variability, between  \textit{weather} (up to decadal time-scales) and \textit{climate} (above one thousand years). This gap indicates the presence of dissipative processes that act as a barrier between the fast and the slow dynamics. It is therefore reasonable to apply the slaving principle. 
In relation to this, Huybers and Curry recently published a composite spectral estimate of temperature variations ranging from sub-daily to Milankovitch time scales \cite{Huybers06aa}. No gap is evident, but Huybers and Curry identify a change in the power-law exponent of the spectral background: Signal energy decays faster with frequency at the above the century time scale than below. 
They interpret this as an indice that  the effective dissipation time scale is effectively larger above the century that below, and, therefore, that the dynamics of slow and fast climatic oscillations are at least  partly decoupled. 

We now enunciate the three differential equations of SM91.

The ice-mass balance is the result of the contribution of four terms: a drift, a term inversely proportional to the deviation of the mean global temperature compared to today ($\tau$),  a relaxation term, and a stochastic forcing representing ``\textit{all aperiodic phenomena not adequately parameterised by the first three terms}'':
\begin{equation}
\frac{dI}{dt}=\varphi_1 - \varphi_2 \bar{\tau} - \varphi_3 I + \mathcal W_I(t).
\label{dIdt}
\end{equation}
According to the slaving principle, $\tau$ is in thermal equilibrium with the slow variables $\{I,\mu,\theta\}$ and its mean may therefore be estimated as a function of the latter:
\begin{equation} 
\bar{\tau}= \bar{\tau}_\tau(I) +  \bar{\tau}_\mu(\mu) +  \bar{\tau}_\theta(\theta) +  \bar{\tau}_R(R),
\label{tau}
\end{equation}
where  $\bar{\tau}_x(x)$ is the contribution variation of $x$ compared to a reference state, to $\bar{\tau}$ keeping the other slow variables or forcing constant. $R$ designates the astronomical forcing (Saltzman used incoming insolation at 65$^o$ N at summer solstice).  The different terms $\bar{\tau}(.)$ were replaced by linear approximations, the coefficients of which were estimated from general circulation model experiments.

The CO$_2$ equation includes the effects of ocean outgassing as temperature increases, a forcing term representing the net balance of CO$_2$ injected in the atmosphere minus that eliminated by silicate weathering,  a non-linear dissipative term and a stochastic forcing:

\begin{eqnarray}
\frac{d\mu}{dt}&=&\beta_0 - \beta_{\theta} \theta + F_\mu - K_\mu \mu + \mathcal{W}_\mu,  \label{dmudt} \\
\textrm{with } K_\mu &=& \beta_1 - \beta_2 \mu + \beta_3 \mu^2. \nonumber
\end{eqnarray}
The dissipative term ($K_\mu \mu$) is a so-called Landau form and its injection into the CO$_2$ equation is intentional  to cause instability in the system. In an earlier paper (e.g. \cite{saltzman88}), Saltzman and Maash attempted to justify similar forms for the CO$_2$ equation on a reductionist basis: each term of the equation was identified  to specific, quantifiable mechanisms like
effect of sea-ice cover on the exchanges of CO$_2$ between the ocean and the atmosphere or that of the ocean circulation on nutrient pumping. It is telling that Saltzman and Maash gradually dropped and add terms to this equation (compare \cite{saltzman88,saltzman90sm,saltzman91sm})  to arrive at the above formulation by which they essentially posits a carbon cycle instability without explicitly caring about which mechanisms that cause it.

The deep-ocean temperature simply assumes a negative dependency on ice volume with a dissipative relaxation term:
\begin{equation}
\frac{d\theta}{dt}=\gamma_1 - \gamma_2 I -\gamma_3\theta + \mathcal{W}_\theta
\label{dthetadt}
\end{equation}

The carbon cycle forcing term $F_\mu$ is assumed to vary slowly at the scale of Quaternary oscillations. It may therefore be considered to be constant and its value is estimated assuming that 
that the associated equilibrium is achieved for a CO$_2$ concentration of 253 ppmv.  We shall note $\{I_0,\mu_0,\theta_0\}$ the point of the central manifold corresponding to that equilibrium, and $\{I',\mu',\theta'\}$  the departure from it. Further constraints are imposed by semi-empirical knowledge on the relaxation times of ice sheet mass balance (10,000 years) and deep-ocean temperature (4,000 years). %This leaves us with us with six parameters for the system determining the trajectory of  $\{I',\mu',\theta'\}$. 

\begin{figure}[t]
\includegraphics[width=\columnwidth]{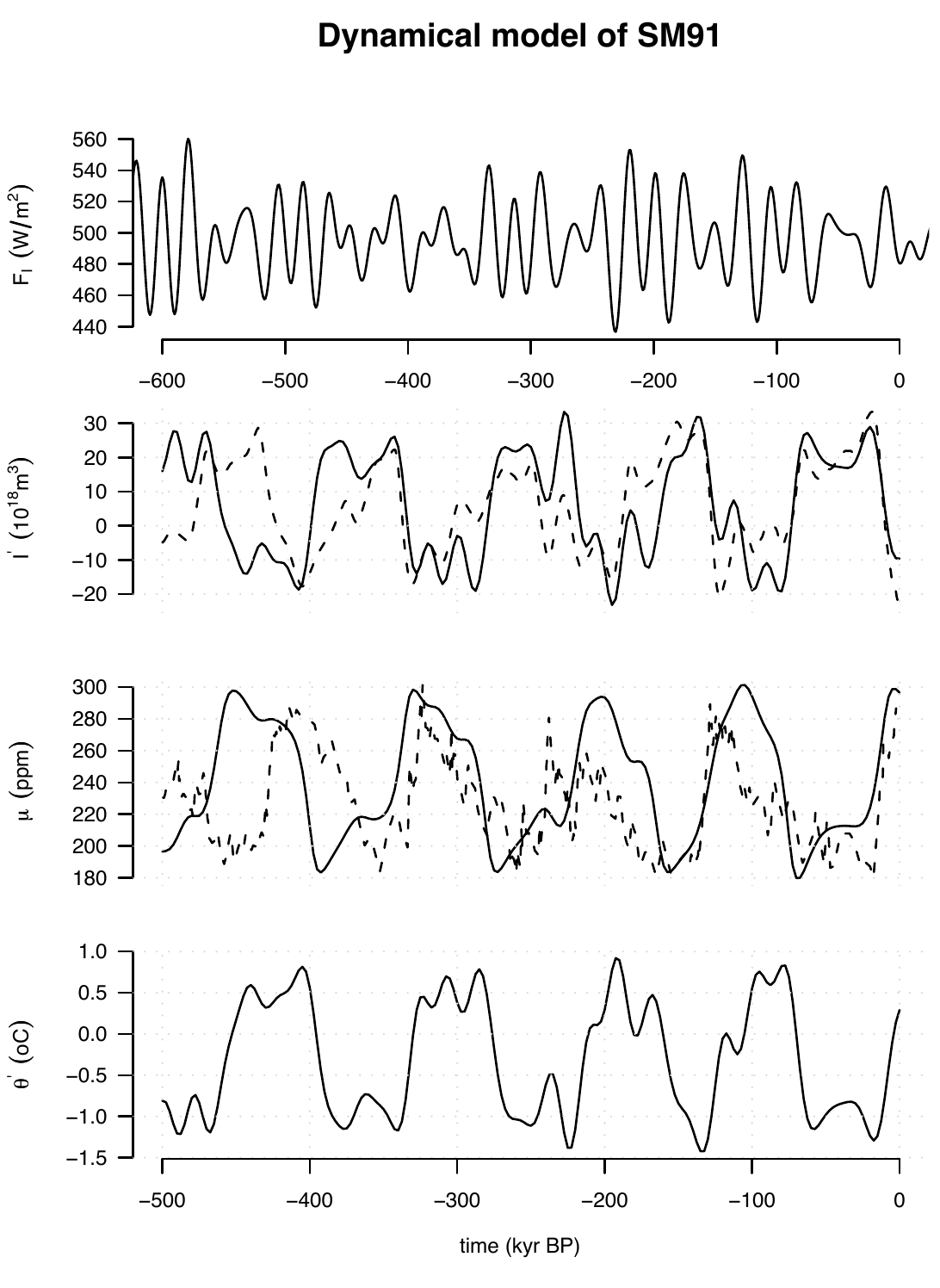}
\caption{Response of the palaeoclimate model of Saltzman and Maasch (1991) \cite{saltzman91sm}. Shown are the insolation forcing, taken as the summer solstice incoming solar radiation at 65$^\circ$ N after \cite{berger78}; the ice volume anomaly (full), overlain with the SPECMAP planctonic $\delta^{18}O_c$ stack \cite{imbrie84} (dashed), the CO$_2$ atmospheric concentration, overlain with the Antarctic ice core data from Vostok and EPICA \cite{petit99,siegenthaler05}, and finally deep-ocean temperature. Note that $I'$ and $\theta'$ are anomalies to the tectonic average. A similar figure was shown in the original article by Saltzman and Maasch}
\label{fig:Saltzman15}
\end{figure}

Saltzman and Maasch explored the different  solution regimes of this system \cite{MAASCH90aa,saltzman91sm} and they observed that climate trajectories converged to a limit cycle characterised by saw-tooth-shaped oscillations for a realistic range of parameters. When the model is further forced by the astronomical forcing, the uncertainty left on the empirical parameters of equations (\ref{dIdt} -- \ref{dthetadt}) provides the freedom to obtain very convincing solutions for the variations in ice volume and CO$_2$ during the late Quaternary. Figure \ref{fig:Saltzman15} reproduces the original solution  \cite{saltzman91sm}, using the parameters published at the time. As in the original publication, the solution is compared with  Imbrie's $\delta^{18}$O-stack \cite{imbrie84} interpreted as a proxy for ice volume, and CO$_2$ record extracted from the Vostok and EPICA(Antarctica) ice cores \cite{petit99,siegenthaler05}. 

\begin{figure*}[t]
\begin{minipage}[t]{0.62\textwidth}
\includegraphics[width=\textwidth]{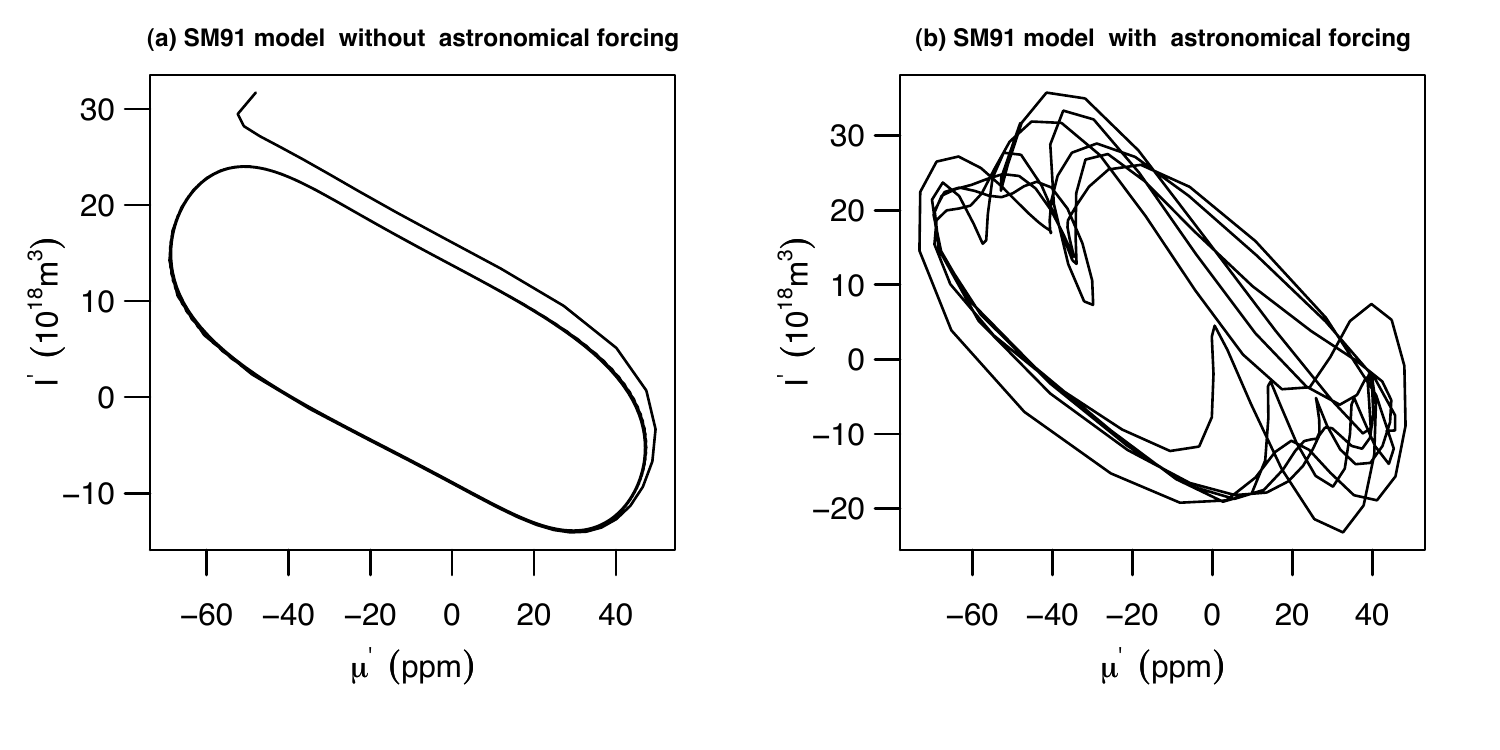} 
\end{minipage}
\hfill
\begin{minipage}[b]{0.30\textwidth}
\caption{Phase-space diagrams of trajectories simulated with the SM91 model, using standard parameters. The model exhibits a limit cycle in absence of external forcing, with a trajectory that resembles those obtained with data (Figure \ref{fig:phase}). The astronomical forcing adds a number of degrees of freedom that complicates the appearance of the phase diagram.}
\end{minipage}
\label{fig:traj}
\end{figure*}

Limit-cycle solutions in SM91 (Figure \ref{fig:traj}) owe their existence to  cubic terms in the CO$_2$ equation. In fact, all parameters being constant, a limit-cycle occurs only for certain carefully chosen values of $\mu_0$, which led Saltzman to conclude that the \textit{cause} of glacial-interglacial oscillations is not the astronomical forcing (a linear view of causality) but rather the gradual draw-down of $\mu_0$ at the tectonic time scale that permitted the transition between a stable regime to a limit-cycle via a Hopf bifurcation.  According to this approach, astronomical forcing controls in part the timing of terminations by a phase-locking process, but terminations essentially occur because negative feedbacks associated to the carbon cycle become dominant at low CO$_2$ concentration and eject the system back towards the opposite region of its phase space.

%Having this in mind, we can see that even though SM91 seems quite ``tunable'', it is falsifiable because the presence of linear terms in the sole CO$_2$ equation imply that the increase in CO$_2$ leads the decrease in ice volume. In fact, the most recent evidences suggest that it is not the case. 

\section{The Bayesian inference process}
%In adopting a model like SM91 we explicitly assume the two decisions outlined in section \ref{remarks}.
Approaches founded on low-order dynamical systems are regularly suspected of being tautological: what can you learn from a model if you tuned it to match observations?  There is no doubt that the empirical content of any model --- i.e., its capacity of being in conflict with observations --- has to be assessed with the utmost care. 
Several authors have in particular insisted on the difficulty of   finding discriminating tests for models with similar dynamical characteristics but built on different interpretations of the climate system's functioning \cite{Roe99aa,tziperman06pacing}. It is therefore challenging but important to identify and design powerful tests for such models.

Nevertheless, it has to be appreciated that the risk of tautology is also present in the most sophisticated general circulation models of the atmosphere and ocean. Once assembled, these models are ``tuned'' to capture major and global characteristics of climate such as the overturning cell or the global mean temperature (e.g.: \cite{jones05famoustuning}). This ``tuning'' is an effective way of incorporating macroscopic information in the model, and this information  can no longer said to be ``predicted''. 

Statistical decision theory allows us to address, at least partly, these difficult problems. We will concentrate on one branch of it: Bayesian inference.  The paradigm of Bayesian inference finds its roots in early works by Bayes, Laplace and Bernouilli who were looking for ways of augmenting their knowledge of certain quantities such as initial conditions or parameters, by means of observations \cite{Jaynes79aa}. Rougier (2007) \cite{Rougier2007Probabilistic-i} explains how Bayesian inference methods may be applied to the problem of climate prediction. His conclusions are summarised hereafter, but adapted were relevant to the problems posed by palaeoclimate time-series analysis. Compared to  Rougier (2007), we more explicitly consider here the fact that climate is a dynamical body, whose evolution has to be predicted by means of time-differential equations.

Before embarking on the mathematical details, it is useful to recall two aspects inherent to complex system modelling introduced in section \ref{remarks}. The first one is that by focussing on certain modes of climate variability we ignore a large body of information, such as its synoptic variability and, for example, the occurrence of a particular volcanic eruption at any particular moment. This ignorance causes prediction errors that we have to \textrm{parameterise}, typically as a stochastic forcing or error (we will here neglect the epistemological distinction between stochastic forcing and error). 
Model \textit{validation} consists in verifying that the model assumptions are compatible with observations. A crucial but often forgotten point is that the validation tests depends on the judgements we will have made about the probability distribution of the model error: 
if we considered that the model error could take any value, the model would always be compatible with observations, but it would also be useless.

The second aspect of complex system modelling is that we accept to consider information that it is not immediately deduced from our knowledge of microscopical interactions. In the case we are busy with, these extra statements take the form of conjectures about the mathematical expressions of carbon, ocean and ice-sheet feedbacks, which are \textit{calibrated} by reference to observations.

Our purpose is to formalise as rigorously as possible the validation and calibration processes. To this end, let us denote $y(t)$ a vector describing the state of climate at a given time $t$. We further notate symbolically ${y}$ the climate evolution over a given time interval not necessarily restricted to the observable past. It is useful to distinguish notationally the variable ${Y}$, which may \textit{a priori} take any value in a given space, from its realisation ${y}$. The exact value of ${y}$ is never known because any measurement or prediction is affected by errors, but the fact of positing the existence and attaching a meaning to $y$ enables us to structure and justify our judgements. 

Palaeoclimatologists attempt to retrieve information  on ${y}$ by taking measurements in a palaeoenvironmental record. Let ${z}$ be a series of such observations like, for example, delta-Deuterium of ice in an Antarctic ice core sampled at certain depths. They estimate that ${z}$ is conditionally dependent on ${y}$, which one may write as :
 \begin{equation} \xymatrix{y \ar^p[r] & z} \label{yarrowz} \end{equation}
This means that their expectation  on $z$ depends on their  knowledge of $y$. This expectation can be quantified by means of a probability density function for $Z$, thought of as a function of $z$: 

\begin{equation}
P(Z=z|Y=y,p) 
\label{zgiveny}
\end{equation}

Building an expression for (\ref{zgiveny}) requires to formulate a number of assumptions forming a \textit{climate proxy model} that we have symbolically denoted $p$. In practice, it may be preferable to decompose this model into a chronological chain of nested processes, each bearing uncertainties: effect of climate on the hydrological cycle, isotopic fractionation, accumulation of ice, preservation of the signal in the core, drilling and actual measurement.  The more there are uncertainties, the wider $P(Z=z|Y=y,p)$ will be. 

Bayesian inversion then indicates us how $z$ is informative on $y$ :

\begin{equation}
P(Y=y|Z=z,p)=\frac{P(Z=z|Y=y) \, P(Y=y)}{P(Z=z|p)}
\label{binversion}
\end{equation}
This equation bears important lessons. First, updating an estimate of $y$ on the basis of observations requires to have some prior judgement expressed in the form $P(Y=y)$. This important question will be kept aside a moment. Second, the denominator at the right-hand-side is independent on $y$. It represents a marginal likelihood, which may be thought of as a point-estimate of a predictive distribution of $Z$ given our prior judgement on $y$ \textit{along with} the assumptions contained in $p$. In practice it is evaluated as:

\begin{equation}
P(Z=z|p)=\int P(Z=z|Y=y,p) P(Y=y)\,dy
\label{validation}
\end{equation}

The validation of $p$ consists in determining if $P(Z=z|p)$ lies in the tails of its distribution. 
The presence of an observation in the tails of its predictive distribution means that it was \textit{little likely} to occur according to the theory expressed in $p$. Such an outcome will incline us to \textit{confidently} reject the theory in the same way that one rejects a null-hypothesis in classical statistics tests.  
This is easily diagnosed in the case where $z$ is a scalar, in which case it may be checked if the marginal probability $P(Z<z|p)$ is not too close to zero or one. 
\begin{equation}
P(Z<z|p)=\int P(Z<z|Y=y,p) P(Y=y)\,dy
\label{validation_upper}
\end{equation}
In practice, $z$ is often highly dimensional and its predictive distribution may be particularly intricate, especially in chaotic dynamical systems. 

At present, it is useful to split $y$ into its ``past'' ($y_p$) and ``future'' ($y_f$) components. If the past is known, the record content is obviously independent on the future, i.e. :
\begin{equation}
P(Z=z|Y_p=y_p,Y_f=y_f) = P(Z=z|Y_p=y_p).
\label{pastfuture}
\end{equation}
\textbf{(\ref{binversion}) and (\ref{pastfuture}) tell us that in absence of any additional assumption, past observations are not informative on the future.} Predicting climate requires to assume a certain dynamical structure to climate evolution to link $y_f$ to $y_h$. This is the role of the climate model. It is in principle always possible to formulate this model in terms of first-order differential stochastic equations if the climate state $y(t)$ is suitably defined. Climate time-series are in this case Markovian:  Given climate at any time $t_0$, the probability density function of climate at time $t_1$ may be estimated and written:
\begin{equation}
P(y(t_1)|y(t_0),c,A=a)
\label{climate model}
\end{equation}
where we distinguish the ensemble of model equations (symbolically denoted $c$) from their parameters, gathered into a single vector variable denoted $A$. 
More generally, the model allows us to estimate the probability density of any climate time-series, which we shall write:
\begin{equation} \xymatrix
{ 
  & y(t_0) \ar^c[d] \\
  a \ar^c[r]&  y
} \label{aarrowy} \end{equation}
 
The model makes it thus possible to build a predictive distribution function for $y$ given any prior estimate of the possible values of $a$ and $y(0)$. 
 
The \textit{climate} and \textit{climate proxy} models may then be combined to form a Bayesian network :
\begin{equation}
\xymatrix{ 
  & y(t_0) \ar^c[d] \\
  a \ar^c[r] \ar^c[dr] &  y_h \ar^p[r] \ar^c[d] & z \\ 
             &  y_p     \\ 
  }
\label{network}
\end{equation}

\textit{Solving} the network means to find the joint distribution of $a$, $y(t_0)$, $y_h$, $y_p$ and $z$ compatible with all the constraints expressed in $p$ and $c$ (e.g. : \cite{Koch07aa}, pp. 167 and onwards).  Keeping in mind that the arrows may be ``inverted'' by application of Bayes' theorem, it appears that there are two routes by which $z$ constrains $y_p$ : via $y_h$ (that is, constraining the initial conditions to be input to the model forecast of the future), and more indirectly via $a$. In the latter route, all observations concur to constrain a distribution of the model parameters that is compatible with both the model structure and the data.

Two more remarks. First, (\ref{network}) shows that the climate model has solved the problem of finding a prior to $y$ (it is provided by the model), but this is at the price of having to find a prior for parameters $a$.  It may happen that one parameter has no clearly identified physical meaning (like $\beta_4$ in eq. (\ref{dmudt})) and we would like to express our total ignorance about it, except for the fact that it is positive. It happens that there is no definitive solution to the problem of formulating a totally ignorant prior. However, if the observations are very informative, the posterior distribution of $a$ is expected to depend little on its prior.

The second remark is about the marginal likelihood, that is, our assessment of the plausible character of observations $z$ given the structural assumptions in models $p$ and $c$ along with the prior on $a$.  It is crucial to be clear about what is being tested. For example, one may be content to assess the position of $z$ thought of  as a $n$-dimensional vector ($n$ is the number of observations) in the manifold of likely $Z$ values given the prior on $a$. This test takes for granted that the stochastic error is effectively white-noise distributed. This being said, it may be useful to effectively test the white-noise character of the model error, typically example by estimating the likelihood of the lagged-correlation coefficients of the stochastic error.
Lagged-correlation coefficients significantly different than zero almost surely indicate that there is information in the stochastic error terms.  This  would prove that the model is incomplete in the sense that its predictive ability can almost surely be improved.

%With this remark in mind we can come back to the way of justifying the number of equations needed to model glacial-interglacial cycles. We remember that Saltzman reduces the number of prognostic equations needed by time-``averaging'' atmospheric and oceanic processes on decadal time-scales and less. He justifies this manoeuvre by the existence of a spectral gap between weather and climate. Unfortunately, we have seen that this argument is not so solid in the light of actual data analysis.
%Rather than seeking for a logical \textit{conclusion} that glacial-interglacial dynamics can be skillfully predicted by a low-order system with stochastic forcing, it might be more purposeful to determine whether there is a \textit{contradiction} between this statement and the various bits of information that are at our disposal, including observation and theory. Our opinion is that such a contradiction is still to be found. For example, the 3-degree-of-freedom of Le Treut and Ghil \cite{letreut83} shows a very wide background continuum of frequencies and displays complex phenomena such as non-linear resonance. The size of the dynamical system needed to model Quaternary oscillations remains an open problem and the statistical theory-based approach outlined above will then offer a suitable framework to assess the consistency between theories and data. 

\section{An application  of the particle filter}
\footnote{This section outlines work in progress carried out by the author in close collaboration with Jonathan Rougier, Department of Statistics at the University of Bristol, UK.}
Network (\ref{network}) is an example of combined parameter and time-varying state estimation problem. This kind of problem is highly intractable, 
but statisticians have been looking at ways of finding approximate solutions based on Monte-Carlo simulations.  Here we use an implementation of the \textit{particle filter} developed by Liu and West \cite{lw01}. This is a filter, that is a sequential assimilation method: observations are used to refine parameter distribution estimates as the time-integration of the model progresses.
The reader is referred to the original publication for a fuller discussion of the method and we will briefly summarise here the sequential algorithm.

First we reformulate  (\ref{network}) into a more tractable problem:

\begin{equation}
\xymatrix{ 
  & y(t_0) \ar^c[d] \\
                       &  y_1 \ar^p[r] \ar^c[d] & z_1 \\ 
                       &  y_2 \ar^p[r] \ar^c[d] & z_2 \\ 
  a \ar^c[uur] \ar[ur] \ar[dr]  \ar[ddr] \ar[ddddr]
                       &  \cdots \ar^c[d] \\
                       &  y_m \ar^p[r] \ar^c[d] & z_m \\ 
                       &  y_{m+1} \ar^c[d] \\
                       &  \cdots\ar^c[d]  \\
                       &  y_n              \\ 
  }
\label{pfilter}
\end{equation}

The important difference with   (\ref{network}) is that the observations are bound to individual state vectors. This implies that their dating is certain (they can unambiguously be associated to a climate state at a given time) and that there is no diffusion of the signal within the record. 

The climate model ($c$) is SM91 (\ref{dIdt} -- \ref{dmudt}), the equations of which are summarised hereafter:
\setcounter{equation}{0} \renewcommand{\theequation}{c\arabic{equation}} 
\begin{align}
\frac{dI'}{dt}&=-a_1 [ k_\mu \mu' + k_\theta \theta' + k_R R'(t)] - K_I I'+ \mathcal{W}_I \\
\frac{d\mu'}{dt}&=b_1\mu'-b_2\mu'^2+b_3\mu'^3-b_\theta \theta + \mathcal{W}_\mu \\
\frac{d\theta'}{dt}&=-c_1I' - K_\theta' + \mathcal{W}_\theta
\label{SM91}
\end{align}

The coefficients $a_i$, $b_i$, $c_i$ and $K_i$ are functions of the $\phi_i$, $\beta_i$, $\gamma_i$ determined using the condition that the equations for $\{I',\mu',\theta'\}$ present a fixed-point at $\mathbf{0}$ (i.e.,  $\{I_0,\mu_0,\theta_0\}$ is a long-term, ``tectonic'' equilibrium). 
Coefficients $k_x$ appear in the process of linearising the short-term response and can in principle be estimated with general circulation models. The reader is referred to the original publications for fuller details. 

The climate proxy model ($p$) is very simple. We will use the \textsc{Specmap} stack of planctonic foraminifera to constrain ice volume \cite{imbrie84}, and the Vostok (Antarctica) ice core by Petit et al. (1999) \cite{petit99} for CO$_2$. 

\setcounter{equation}{0} \renewcommand{\theequation}{p\arabic{equation}} 
\begin{eqnarray}
\delta^{18} \mathrm{O}_c &=& \frac{0.71}{  45 \ 10^{18}\  \mathrm{m}^3 } I'+ \mathcal{W}_\delta \label{p1} \\
{\mathrm{CO}_2}   &=& \mu + \mathcal{W}_\mathrm{CO2}  \label{p2}
\end{eqnarray}

Equation (\ref{p1}) uses the fact that the Imbrie et al. record is expressed in standard deviation units with zero mean, along with the constraint that a total ice melt of $45\ 10^{15}$\ m$^3$ is recorded as a drop of 0.71 (unitless) in Imbrie et al. We therefore neglect the influence of ocean temperature on the record, while this issue is contentious. Errors are parameterised by means of additive stochastic Gaussian white noise with standard deviations of $0.2$ (\ref{p1}) and $20\ \mathrm{ppm}$ (\ref{p2}), respectively. 

The above approximations (neglecting dating uncertainty, in-core diffusion and unduly simple isotope model) will no longer be tenable as this research project develops but they are suitable for a first application of the particle filter algorithm. Consequently, results should be considered with the necessary caution. 

\begin{figure}[t]
\includegraphics[width=\columnwidth]{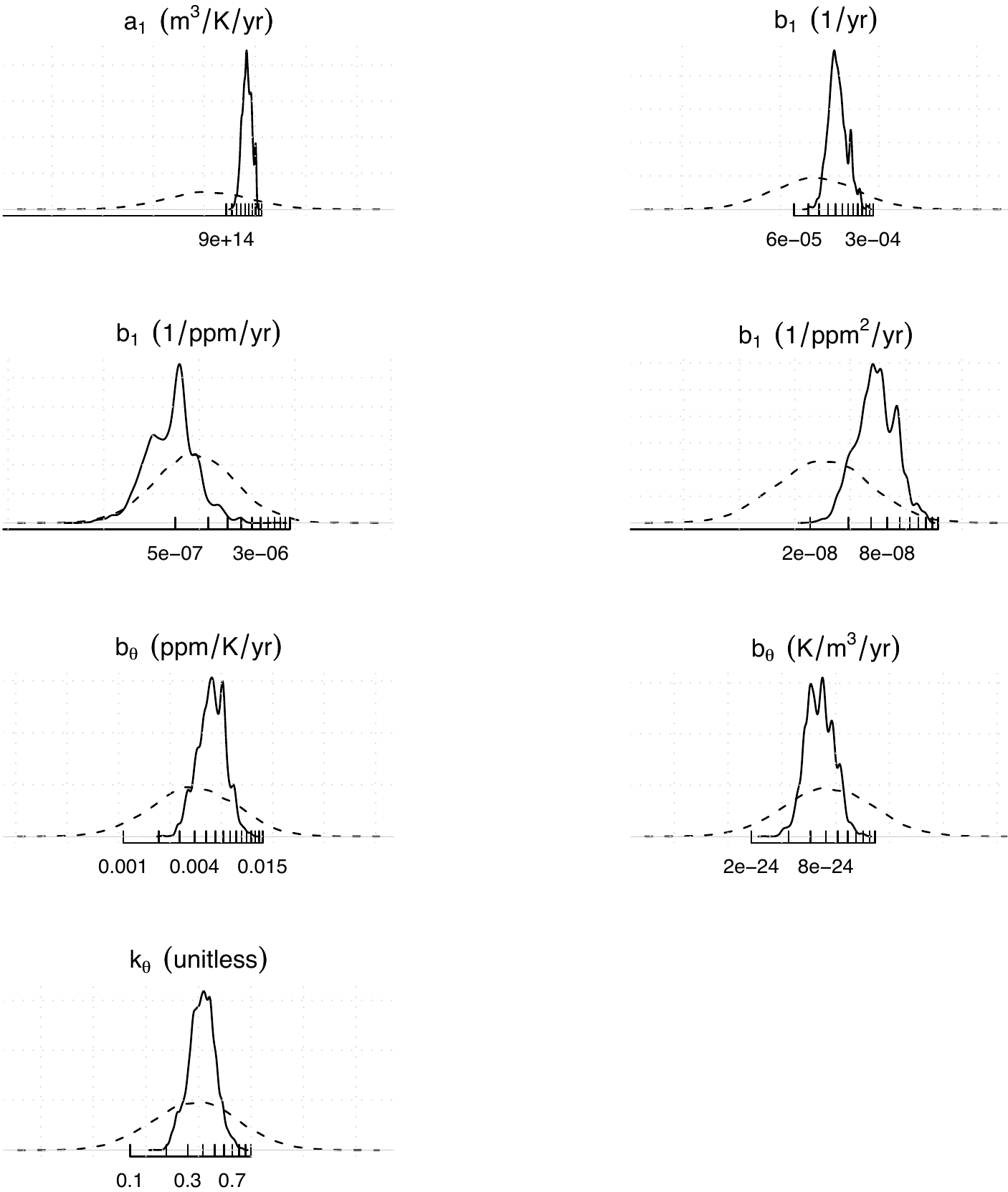}
\caption{Prior (dashed) and posterior (full) density estimates of the parameters allowed to vary in SM91. The filter has been successful in narrowing down the distributions. }
\label{fig:DensityReal}
\end{figure}

We now review the particle-filter algorithm. A particle is essentially a realisation of the state vector (say : $y(t_0)$) associated to a realisation of the parameters ($A=\{\ln (a_i,b_i,c_i)\}$) and a weight ($w$). Ten thousand ($n$) particles are initialised by sampling the prior of $y(t0)$ and $A$.  Prior parameter distribution are log-normal around the values given in SM91 (Figure \ref{fig:DensityReal}). Only the $a_i$, $b_i$, $c_1$ and $k_\theta$ are considered to be uncertain, while the dissipative exchange coefficients $K_I$ and $K_\theta$ as well as the climate sensitivities $k_\mu$ and $k_R$ are assumed to be known (Table \ref{table:fixed}).

All weights are initialised to $1$. The filter then consists of an iterative six-step process. Say we are at time $t$.
\begin{enumerate}
\item \textbf{Propagation}, that is, time-integration of all particles until the time ($t+1$) corresponding to the next available data (either CO$_2$ or $\delta^{18}\mathrm{O}$.
\item \textbf{Shrinkage}. Particles are now dispersed in a region of the $\{Y,A\}$.  This region in \textit{shrunk}, that is, the particles are made closer to each other by a factor $\alpha$.
\item \textbf{Weight estimate} Particle weights are multiplied by the likelihoods $P(Z=z|Y=y_j)$, where $y_j$ is the state of particle $j$, and $z$ is the encountered data.
\item \textbf{Importance resampling based on posterior estimate}. After step 2, some particles may be given a large weight while others only a small one. Particles are therefore resampled in such a way that they all get a similar weight. This implies that some particles are duplicated while others are killed. Particles are now distributed along $k<n$ kernels. \label{step4}
\item \textbf{Resampling of kernels} Each kernel is broken apart into particles with parameters scattered with variance $h^2$.\label{step5}
\item \textbf{Weight update} Particles weights are updated according to their likelihood.
\end{enumerate}
Shrinkage and kernel sampling are artefacts introduced to avoid filter degeneracy.  Liu and West \cite{lw01} note that the estimator is unbiased for $\alpha=(3\delta-1)/2\delta$ and $h^2=1-\alpha^2$. The parameter $\delta$ is called a \textit{discount factor}. It must lie in $]0,1]$ and typically around 0.95 -- 0.99. Here we choose $\delta=0.95$. We found that the parameter disturbance due to the filter dominates any reasonable amount of stochastic error that could be parameterised via $\mathcal{W}$. Therefore, we decided not to account for the model stochastic error noise to gain computing efficiency.

\begin{figure}[t]
\includegraphics[width=\columnwidth]{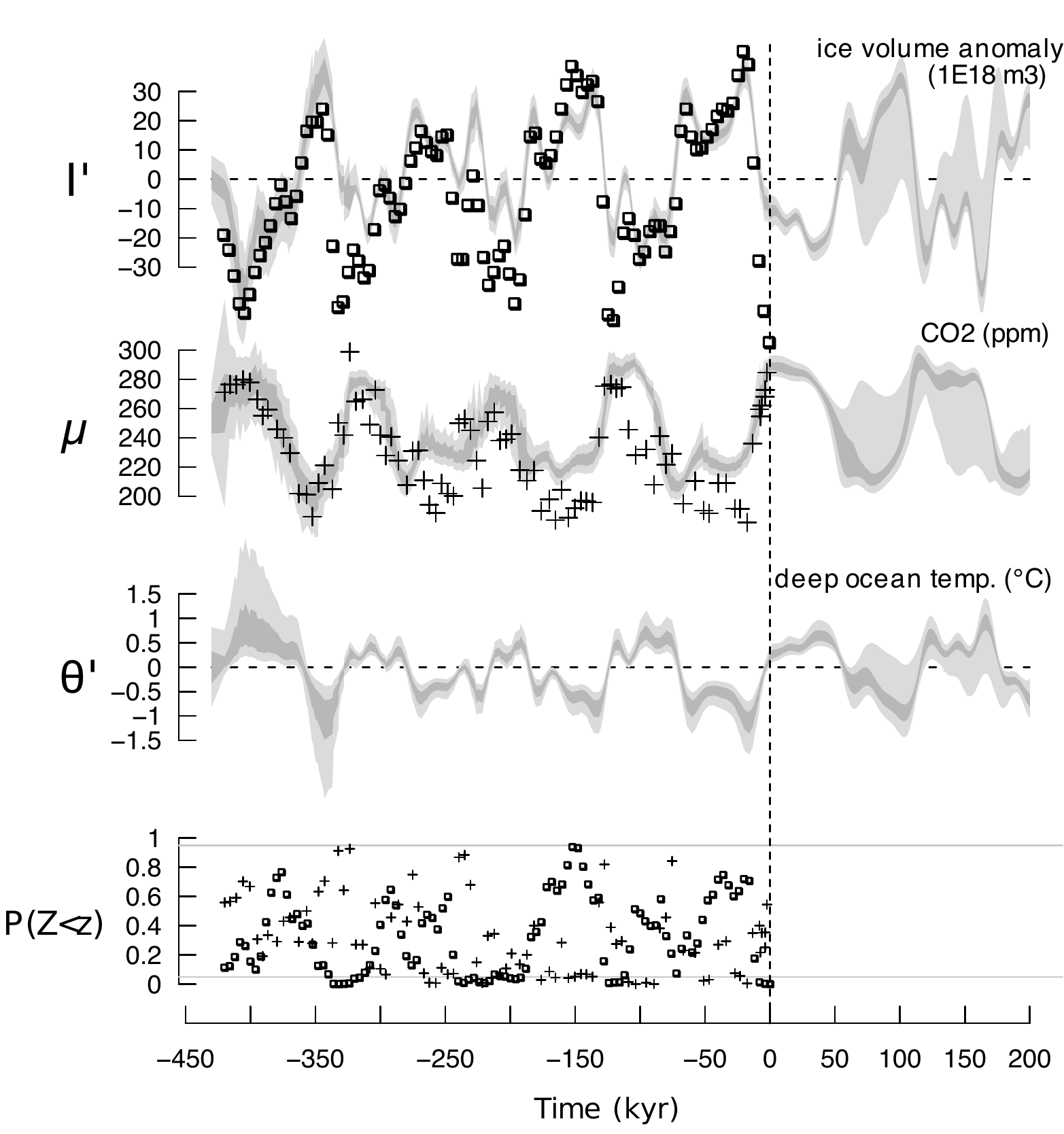}
\caption{Filtered state estimates with the SM91 model constrained by the SPECMAP data (squres), and the Antarctic ice core data (pluses). The state estimates are represented by shades, dark and light gray representing  the $[25^\textrm{th};75^\textrm{th}]$ and $[5^\textrm{th};95^\textrm{th}]$ quantiles of the particle weighted distributions, respectively. The lower graph represents, for each data, the model predictive probability that the data would have been lower than it actually was, given the previous parameter and state estimates. The repetition of probabilities below 0.05 or above 0.95 tend to invalidate the model.}
\label{fig:stateFilterReal}
\end{figure}

Figure \ref{fig:stateFilterReal} summarises the essential features of the particle filter run. It represents, for each prognostic variable, the evolution of the state estimate (shaded) along with the data. The dark and light shades represent the central 50 and 90 \% percentiles of the weighted particle distribution. The filter algorithm updates the parameter estimates as it meets the data (The posterior parameter distributions are compared to the prior on Figure \ref{fig:DensityReal}), which explains why the state estimates become narrowed as time progresses. The dots and pluses are the observation estimates of ice volume and CO$_2$. 
The fourth panel is a first step towards model validation. It display, for each observation, the model predictive probability that this observation was smaller or equal than its value, exactly in the spirit of equation (\ref{validation_upper}). Values too close to zero or one cast doubt on the model. 

It was unexpected that the fit of the state estimates of the ice volume on SPECMAP would be so poor. In fact, the model systematically overestimates ice volume during interglacials and this occurs as soon as CO$_2$ observations are taken into account.  Strictly speaking, the model is invalidated. Where does the problem lie ? 

The most obvious possibility is that we have incompletely modelled the SPECMAP stack. Indeed, we know that water temperature contributes to the $\delta^{18}O_c$ signal but this contribution is missing in the model
(\cite{shackleton00,emiliani92,Adkins02,Bintanja05aa}, the latest reference being another example of data reanalysis).

In spite of this weakness, we will assume that the model estimates of $I'$ give a correct representation of ice volume anomalies around a tectonic-time-scale  average.
Ice-volume levels typical of the last interglacial then correspond to $I'=-15\cdot 10^{18}$ m$^3$ in the model. The model prediction is an immediate but slow decrease in CO$_2$ concentration (Figure \ref{fig:stateFilterReal}) but without glacial inception  before about 50,000 years (this is the Berger and Loutre prediction \cite{Berger2002An-exceptionall}!). The particle filter also tells us that given the information at disposal (the model, the data, and the parameter priors), it is not possible to provide a reliable estimate of the evolution of climate beyond 50,000 years. 

\begin{figure}[t]
\includegraphics[width=\columnwidth]{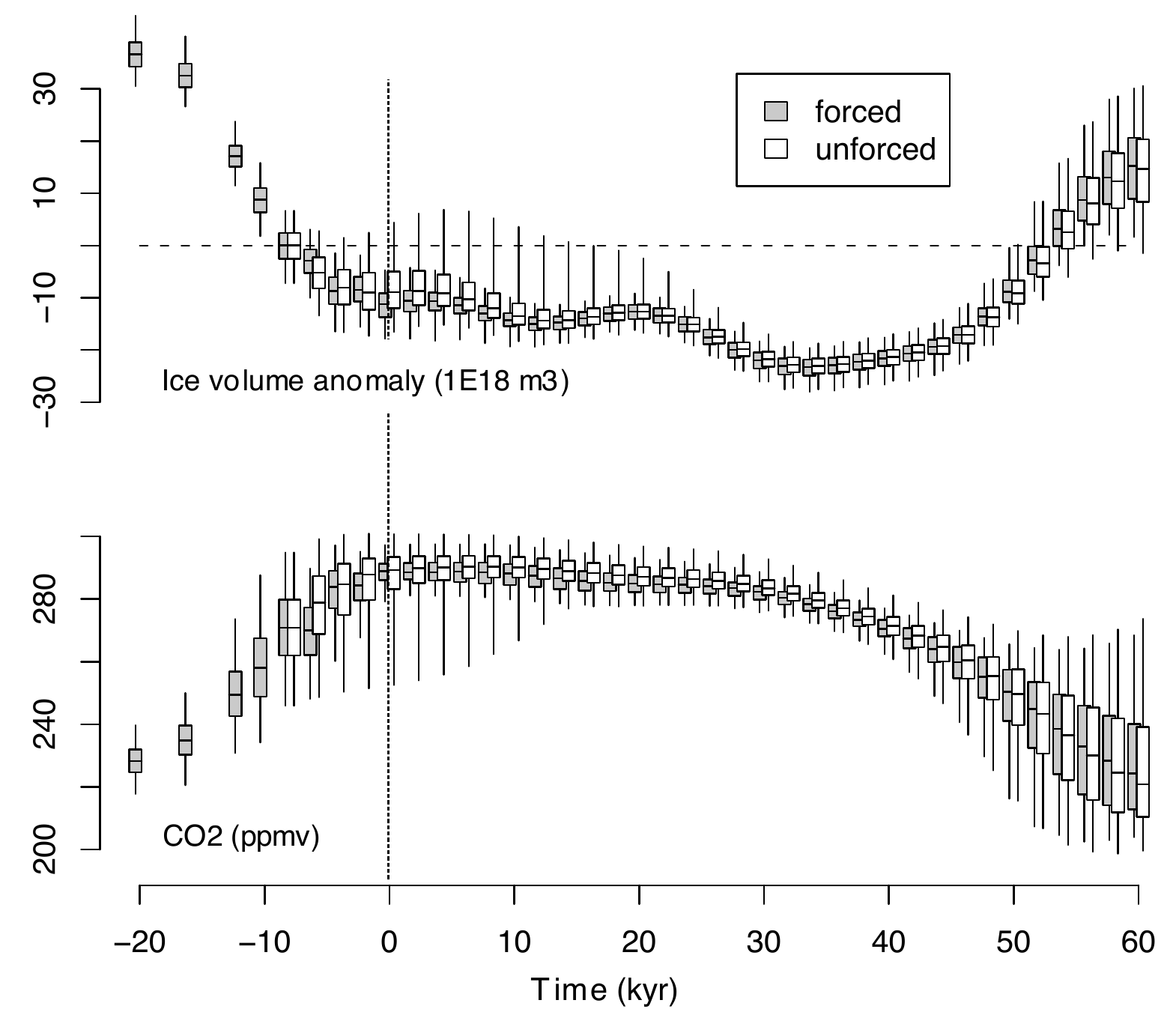}
\caption{State estimate with the SM91 model, given data on CO$_2$ and ice volume between 410 kyr BP and 8 kyr BP \textit{(white)} or 0 lyr BP \textit{(grey)}. The subsequent prediction, with glacial inception in 50 kyr, is little affected by the data between 8 and 0 kyr BP. This is opposed to  Ruddiman's hypothesis}
\label{fig:stateCompReal}

\end{figure}

What about Ruddiman's hypothesis? Ruddiman considers that humans perturbed climate's evolution around 8000 years ago. Therefore, we want to only consider data until that time, and see whether the model prediction differs to the previous one. The experiment was carried out and the results are presented on Figure \ref{fig:stateCompReal}. The grey boxes provide the prediction with data assimilated until 8,000 years ago, and the white ones is the prediction with data assimilated until today. The two predictions are clearly undistinguishable. Contrarily to Ruddiman, our model was therefore not ``surprised'' by the fact that CO$_2$ continued to increase during the last 6,000 years.

\begin{table}
\centering
\begin{tabular}{llll}
\hline
parameter &  fixed value  \\ 
\hline
$k_\mu$   & 0.04 K /  ppm /yr\\
$k_\theta$ & 0.5 1/yr      \\
$k_R$     & 0.08 K /  Wm$^{-2}$/yr \\
$K_I$     & 1.e-4 1/yr \\
$K_\theta$     & 2.5e-4 yr$^{-1}$ \\
\hline
\end{tabular}
\caption{Values of SM91 fixed parameters used both in the original publications and in the present article}
\label{table:fixed}
\end{table}

\section{Conclusion} 
Behind this paper is the message that climate modelling is not and should not be a mere technological question. Of course, general circulation models skillfully predict many complicated aspects of atmosphere and ocean dynamics; in that sense they are important and useful. Yet, they are but one aspect of the theoretical construct that underlies state-of-the-art knowledge of the climate system. Important questions are \textit{how} we validate and \textit{calibrate} climate models to provide the most informed predictions on climate change.

Palaeoclimates offer a premium playground to test the paradigms of complex system theory. We have been insistent on the fact that palaeoclimate theory must rely on two pillars of modern applied mathematics: dynamical system theory and statistical decision theory. Along with the fact that palaeoclimate data have to be interpreted and retrieved by skillful field scientists, their analysis turns to be a truly multidisciplinary experience. The exceptionnally difficult challenges so posed are definitevely at the frontier of knowledge.

\section{Short Biographic note about the author}
Michel Crucifix graduated in Physics in Namur University (1998), followed by a Master in Physics (1999) and a doctorate in Sciences (2002) in Louvain-la-Neuve (Belgium). The subject of his thesis was "modelling glacial-interglacial climates over the last glacial-interglacial cycle". The origin and mechanisms of glacial-interglacial cycles that have paced the Earth over the last million years has remained his favourite topic throughout his young career. He was appointed permanent research scientist at the Meteorological Office Hadley Centre (UK) in 2002 and then came back to Belgium, his home country, as a research associate with the Belgian national fund of scientific research, and lecturer in two universities (Louvain-la-Neuve and Namur). %Michel is author or co-author of about 35 peer-reviewed articles, some of which were produced in the context of broad international and multidisciplinary collaborations, like the Palaeoclimate Modelling Intercomparison Project. He is now most interested in the application of dynamical system theory and Bayesian inference formalism to climate modelling. 

\section{Acknowledgements}
I would like to  thank the organisers of the ``Complexity'' workshop -- Heidelberg for their kind invitation. I would like also to thank Jonathan Rougier for his thoughtful comments on an earlier version of this paper.

\bibliography{/Users/crucifix/Documents/BibDesk.bib}

\end{document}